\documentclass[conference]{IEEEtran}

\usepackage{epsfig}
\usepackage{epstopdf}
\usepackage{amssymb}
\usepackage{amsmath}
\usepackage{amsfonts}
\usepackage{array} 
\usepackage{ctable}
\usepackage{booktabs}
\usepackage{subfigure}
\usepackage{url}

\def\hlinewd#1{%
\noalign{\ifnum0=`}\fi\hrule \@height #1 %
\futurelet\reserved@a\@xhline}

\DeclareFontFamily{OT1}{pzc}{}
\DeclareFontShape{OT1}{pzc}{m}{it}{<-> s * [1.10] pzcmi7t}{}
\DeclareMathAlphabet{\mathpzc}{OT1}{pzc}{m}{it}

\newcommand{\norm}[1]{\lVert#1\rVert}

\ifCLASSINFOpdf
\else
\fi

\begin{document}
\title{Using Location-Based Social Networks to Validate Human Mobility and Relationships Models}

\author{
		Tommy Nguyen, Boleslaw K. Szymanski\\
		Rensselaer Polytechnic Institute \\
		\{nguyet11, szymab\}@rpi.edu\\
}



%

\maketitle
\begin{abstract}
We propose to use social networking data to validate mobility models for pervasive mobile ad-hoc networks (MANETs) and delay tolerant 
networks (DTNs). The Random Waypoint (RWP) \cite{1208967} and Erdos-Renyi (ER) models have been a popular choice among researchers for generating mobility traces of nodes and relationships between them. Not only RWP and ER are useful in evaluating networking protocols in a simulation environment, but they are also used for theoretical analysis of 
such dynamic networks. However, it has been observed that neither relationships among people nor their movements are random. Instead, human movements frequently contain repeated patterns and friendship is bounded by distance. We used social networking site Gowalla 
to collect, create and validate models of human mobility and relationships for analysis and evaluations of applications in opportunistic networks such as sensor 
networks and transportation models in civil engineering. In doing so, we hope to provide more human-like movements and social 
relationship models to researchers to study problems in complex and mobile networks. 
\end{abstract}

\section{Introduction}
Mobile networks are dynamic networks that are created by connections between users' devices. These devices are small enough for a 
human being to carry around. For example, cellphones allow us to keep in contact and handheld-transceivers allow soldiers to 
communicate with their commander. There are two important features of mobile networks. One is mobility of the nodes resulting from the 
intrinsic nature of humans that compels them to travel with their devices from one location to another. Another important feature is that 
direct communication between any two devices is only possible when they are  within transmission range of each other. These two 
features make such networks highly dynamic in terms of their connectivity and strongly dependent on human mobility patterns.

The performance of a mobile network depends on a number of factors such as the routing protocol, mac protocol, 
and topology. Researchers can make changes to protocols to optimize their performance, but they cannot control the 
topology or the mobility. For instance, the topology of a military network is dictated by who is allowed to communicate with whom. 
Only soldiers are allowed to be part of the military network. Everyone else is denied access by default. Similarly, mobility is impacted by the locations of popular interests for human carriers of the devices. 

As pointed out in \cite{wang:mobility}, mobility and social relationships are important not just for mobile networks. Public health, city planning, traffic engineering and economic forecasting can benefit from the knowledge of statistical patterns that characterize the trajectories of movements in humans during their daily activities. For instance, health organizations may want to be able to predict the spread of contagious diseases while traffic engineers may want to 
model a system where travellers can use a combination of bikes, buses, and subways to get from one location to another. 
Using real and large scale data to understand human mobility is critical to such applications

To understand mobility, researchers have resorted to synthetically created mobility traces and social 
relationships. For instance, the random waypoint (RWP) \cite{boldrini2007users} has become a popular means to researchers for providing mobility model for
a plethora of applications. In RWP, each node moves independently from 
each other. Each node starts at a random location and moves to a randomly chosen location with a constant speed. Once the node reaches 
its destination, it pauses for some random time. The stationary distribution of movements can be approximated by $f(x,y)\sim f(x)f(y) = \frac{9} {16x_{m}^{3}y_{m}^{3}}(x^{2} - x_{m}^{2})(y^{2} - y_{m}^{2})$ for the RWP \cite{Bettstetter03thenode}. Intuitively, if we pick two 
random points on the finite compact area of an euclidean space and draw a straight line connecting them, then it is likely that the 
straight line crosses the center of gravity of the area.  This is because the density of the nodes is not uniformly distributed but it is highest near the area of center of gravity

In graph theory, the Erdos-Renyi (ER) model is often chosen by researchers for generating random graphs. One variant is to permute all possible subgraphs by varying the edges and randomly choosing one subgraph from all possible combinations with equal probability. 

Is it appropriate to use RWP and the ER to model human mobility or applications that depend on it? 
First, we argue that humans do not move randomly from one point to another. Instead, it has been observed that humans move in 
repeated patterns with a bursty behavior \cite{human-mobility}. Second, our intuition tells us that humans do not move 
independently. Friends, colleagues, and family members travel together in groups. These two social properties of human mobility 
violate the essence of RWP and ER. 

While individual mobility is a well studied topic within the last few years \cite{human-mobility}\cite{cho:movement}\cite{wang:mobility}, group mobility is an expanded concept that studies 
how humans move together with friends, family, colleagues, or a group with any other social ties. For many applications, assuming that humans move independently from each other is not realistic. 

Studying group mobility is difficult because of the lack of data recording movements of people. Recently, the popularity of social 
networks resulted in a plethora of applications that let researchers collect massive amount of data on human behavior. Gowalla is a 
location-based social networking provider that allows users to share their geographic location with their friends through their smart 
phones in the process known as ``checking in." Similarly, FourSquare and Google Latitude allow users to share their current 
geographic location with their friends. 

While it is true that other social media like Facebook and Twitter provide as an unintended consequence of people locations through geo-tagged posts and tweets, Gowalla was designed to provide such a mechanism. 

Our objectives are as follows. First, we want to understand the power and limitation of the data available from Gowalla for providing 
insights on how distance limits the possibility of friendship. Second, we want to provide a friendship mobility model by 
using a Markov Model derived from movements of groups of people chosen on the basis of friendship. Third, we want to implement our friendship-based mobility model framework in ns-2 \cite{ns2}, so 
researchers can use our open source code and training datasets to evaluate their applications, like modelling traffic congestion in 
urban areas. And finally, we want to compare and contrast the results of traffic congestion using empirical data on friendship and mobility with the RWP. 

The rest of the paper is organized as follows. In section \ref{sec:data-collection}, we present the methodology for acquiring 
the mobility datasets that we use in this paper. In section \ref{sec:checkin-analysis}, we define some attributes and formulas for 
analyzing the dataset. In section \ref{sec:mobility-generation}, we present a framework for using a Markov Model to generate mobility 
traces. In section \ref{sec:protocol}, we compare the results of network congestion in a MANET by running one set of simulations in ns-2 
with the random waypoint model (RWP) and another set with our friendship-based mobility model that we have named FMM. Before concluding in Section 
\ref{sec:future-work} with a summary and discussion of future work, we present the existing literature of replicating human 
mobility in section \ref{sec:literature}. 

\section{Data Acquisition}
\label{sec:data-collection}
By using the Gowalla's API\footnote{Unfortunately, Gowalla has been purchased by Facebook and is no longer operational.}, we were able to retrieve 391,223 users with public profiles (friends and checkins) from mid September in 
2011 to late October of that year. Some of the locations at which users in our dataset checked in are listed in Table \ref{checkin-locations}. First, we start with a user randomly chosen and process all the public information available about 
that user. Second, we store all id's of the user's friends and put them into a processing queue in a FIFO order. Then we retrieve the next 
user from the queue and repeat the process. Therefore, we crawled Gowalla breadth-first, a standard technique in the social networking 
literature often referred to as Breadth First Search (BFS) sampling.  As shown in Table \ref{data-summary}, the users accumulated a total of around 26 million checkins and 8 million friendship links. The geographical spread of the checkins is shown in Fig. 1.  


In \cite{DBLP:kurant}, Kurant et al. argued that BFS sampling is highly biased toward nodes with high degree because such nodes are  more likely to be sampled than nodes with low degree. Since we do not know the exact population of the users on Gowalla, the 
best we could do was to estimate the size of it by using statistical analysis. Using collision counting proposed by 
\cite{Razafindralambo_2006}, we estimated that the population of Gowalla during the period of our BFS sampling was 
500,000. Hence, this number gives us a sense that 
even though BFS is biased towards high degree nodes, the population of Gowalla was small enough for our purposes. 

To summarize the collision counting algorithm for estimating the population size introduced in \cite{Razafindralambo_2006}, let $r$ be the number of samples taken independently from an empirical graph $G=(V, E)$. A sample of $G$ is a subgraph $H$ where the vertices in $H$ are chosen with their respective probabilities from $G$. The probability of a node $v_{i}$ being chosen for $H$ is $d(v_{i}) / D$ where $D = \sum{d(v_{j})}$ $\forall j \in V$ and $d(v_{j})$ is the degree of node $j$. Let $I$ be the number identical nodes being sampled across $r$ subgraphs and $D'$ be the sum of degrees for each sampled node across r subgraphs. Using expectation, the size of the network is approximated as

\begin{equation}
	\label{networksize}
		n = {r \choose 2} \frac{\mathbb{E}[D'] * \mathbb{E}[D'^{-1}]}{r^{2} * \mathbb{E}[I]} \approx \frac{\mathbb{E}[D']\mathbb{E}[D'^{-1}]} {2\mathbb{E}[I]} 
\end{equation}

There are limitations of our dataset and analysis that are worth mentioning. Due to the discretization within the dataset, the route 
from one checkin to another is a straight line because we do not know how a user goes from one place to another. In reality, road and 
building structures enforce non-linear trajectory of human mobility. For privacy 
considerations, we do not have access to the checkins that the users do not share. Therefore, the time interval between two publicly 
available and consecutive checkins can be high. 

Mislove et al. \cite{mislove-2011-twitter} mentioned that the 
population of users who tweet on Twitter is unbalanced. Therefore, we believe that the users who checks in on Gowalla do not make a
representative sample of the entireregarding colors
Fig. 1 and 2 are fine

Fig 3a, please make friendship (crosses) black, and non-friends (dots) gray or green
Fig 3b, please make FMM moves black and RWP gray or green dotted
Fig 3c, change friendship line from red to green
Fig 4a and b are fine
Fig 4c make congestion for FMM black and for RWP green or gray population (e.g., income level, age and gender impact the probability of being such a user). Last but not least, we believed that popular locations such as coffee-houses, movie theatres and airports are checked in more often than private locations.

\begin{table}
	\caption{Selected Locations that users have reported}
	\center
	\label{checkin-locations}
	\begin{tabular}{|c|ccc|r|}
		\hline
			Location & Lat & Lng & Occurrences \\
		\hline
			Austin-Bergstrom Airport	& 30.20155	& -97.66712 & 21,000 \\
			Apple Headquarter	& 37.33188 & -122.02963	& 2,200 \\ 
			Sleeping Beauty Castle	& 33.81335 & 117.91870 & 1,000 \\
			Odd Duck Farm to Trailer & 30.25414 & -97.76231 & 1,000 \\
			Boston University & 42.35115 & -71.10767	& 200 \\
			15 Central Park West Condo & 40.77056 & -73.98146	& 100 \\
		\hline
	\end{tabular}
\end{table}

The average number of friends per user in the Gowalla dataset is 11 with a standard deviation of 67, which is bounded by Dunbar's number \cite{dunbar-num} that suggests humans can cognitively keep at most 150 meaningful relationships. We argue that since privacy and safety are concerns, users are more likely to share their geographic location with someone who they actually know in reality. 



\section{Checkins Analysis}
\label{sec:checkin-analysis}

	\begin{figure}
		\begin{center}
			\includegraphics[width=90mm, height=40mm]{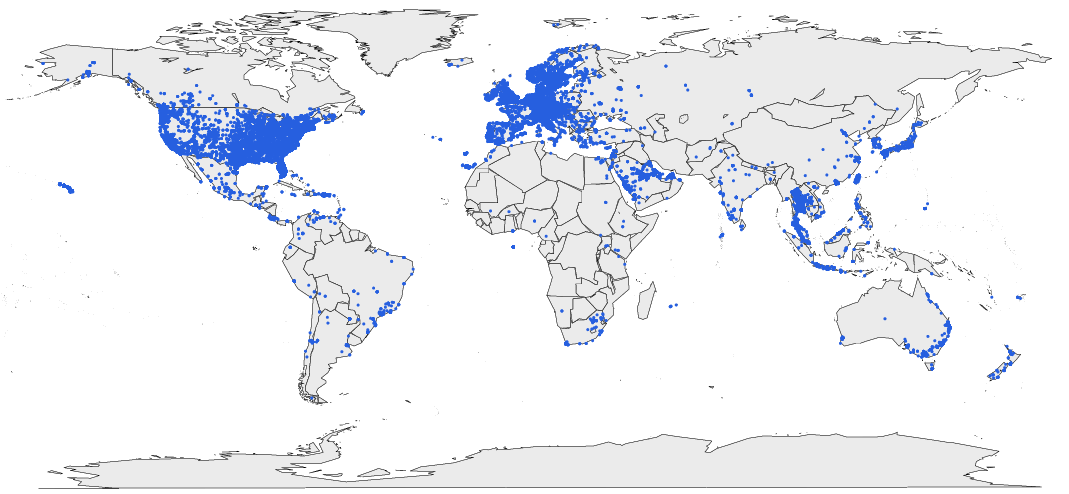}
			\caption{Displaying the geographic location of a randomly selected set containing 100,000 checkins from Gowalla. Notice the dataset is constrained by the economic availability of smartphones and the popularity of Gowalla in a country is correlated to the GDP of this country.}
		\end{center}
		\label{fig:checkin-spread}
	\end{figure}

	\begin{table}
		\caption{Data summary of Gowalla}
		\center
		\label{data-summary}
		\begin{tabular}{|c|ccc|r|}
			\hline
		 		&  $\bar{x}$ & $\sigma_X$  & $\sum_{}$ \\
			\hline
				Users      & $-$ & $-$ & 391,223  \\
				Checkins   & 164.64 & 636.68 & 26,303,580 \\
				Friends    & 11.13 & 67.03 & 2,176,384  \\
				Weekday    & 3.14 & 2.01 & Jan. 21, 2009  \\	
				Distance   & 128.72 & 356.51 & 20,565,644  \\
				Time	   & 6.41 & 13.29 &  - \\
			\hline
		\end{tabular}
	\end{table}

As Table \ref{data-summary} shows, the total number of users in the dataset is 391,223. The average number of checkins for a 
user is 164.64 with a standard deviation of 636.68. The average day of the checkins is 3.14 which represents Wednesday. The 
earliest checkin is on Jan. 21, 2009. The average time interval between two consecutive checkins of a user is 6.41 days with a 
standard deviation of 13.29.

We used the Haversine formula to calculate the shortest distance between any two geographic coordinates. By 
assuming that the Earth is spherical, we calculate the distance by taking the shortest arc between two points on a 
sphere instead of going through the interior. We take the arc instead of a line because for longer routes Earth curvature matters, but 
it is still just an approximation because routes from two given location are not 
necessarily a straight arc in reality due to road structures and traffic. Practically, we could use Google Map to calculate the 
expected distance and time it takes to get from one location to another location.  Unlike Haversine formula, Google Map factors into 
street structures, possible routes, and multiple methods of transportation (bike, bus, car, etc.), which makes it unnecessarily complicated for our purposes. 

The following is the formula for calculating distance between two points $a$ and $b$. $\alpha_i = (\alpha_i^0,\alpha_i^1)$ is a 2-tuple with the first element corresponding to 
latitude and the second element corresponding to longitude. The distance is defined in terms of $\alpha_{a}$ and $\alpha_{b}$ with $r$ being the radius of the Earth. 

	\begin{eqnarray}
		\label{haversine}
			d(\alpha_{a}, \alpha_{b}) = 2r * sin^{-1}(\sqrt{\phi}) \\
			\phi = sin^{2}(\frac{1}{2} \Delta_{lat}) + cos(\alpha_{a}^{0})cos(\alpha_{a}^{0})sin^{2}(\frac{1}{2} \Delta_{lng}) 
	\end{eqnarray}
	
where $\Delta_{lat}=\alpha_a^0-\alpha_b^0$ and $\Delta_{lng}=\alpha_a^1-\alpha_b^1$ are the differences between latitudes and longitudes of the two points.  

The checkin similarity of two users, $i$ and $j$, denoted by $CS(i,j)$ is defined as

	\begin{equation}
		\label{checkin-similarity}
		CS(i,j) = \frac{|C_{i} \cap C_{j}|}{|C_{i} \cup C_{j}|}
	\end{equation}

where $C_{i}$ denotes a set of all checkins of the user $i$. Two checkins of different users are the same, if and only if, they occur within the narrow intervals of time\footnote{Checkins are timestamped.} and space (latitude and longitude). 
Here we want to allow for some difference in data, two checkins within seconds in nearby locations should be treated as the same. Conceptually, the checkin similarity represents how often do these two users occur at 
the same time and place among all their checkins. 

The average distance between two users $i$ and $j$ denoted by $d(i, j)$ is defined as

\begin{equation}
	\label{distance-similarity}
	d(\bar{\alpha_{A}}, \bar{\alpha_{B}}) 
\end{equation}

where $\bar{\alpha_{i}}$ represents the average position of user $i$ with $k$ checkins defined\footnote{Result for people travelling a lot may be misleading, so in the future work we will eliminate users with long range checkins from our analysis.} as: 

\begin{equation}
	\bar{\alpha_{i}} = \frac{1}{k} [\sum\limits_{j=1}^{k} \alpha^{0}_{j}, \sum\limits_{j=1}^{k} \alpha^{1}_{j}]
\end{equation}

Conceptually, we take the average $\bar{lat}$ latitude and $\bar{lng}$ longitude of two users and use the formula (\ref{haversine}) to calculate the average distance between them.

\section{Mobility Generation}
\label{sec:mobility-generation}
\begin{figure}
		\begin{center}
			\includegraphics[scale=0.60]{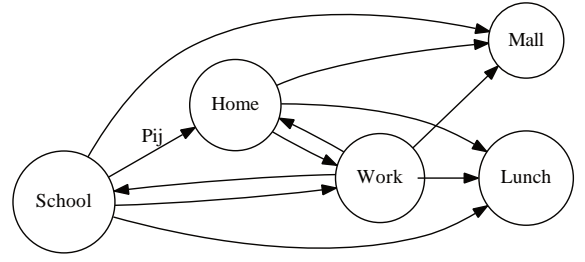} 
			\caption{Generating an empirical Markov Model using checkins. The states represent locations of checkins and the links represent the probability of going from one checkin to another. The probability of going from school to lunch is defined by the datasets as a ratio of the number of times a given user checks in at lunch right after checking in at school to the number of time that user checks in at school.}
			\label{generating-mm}
		\end{center}
	\end{figure}

We propose a following algorithm for generating mobility traces using social networking data from Gowalla or any other location 
based social network. For our Friendship Mobility Model (FMM) using Markov Model as an underpinning, we first randomly select a user from the dataset and include his or her 
friends into the selected group of users. For each user selected, we calculate the patterns of checkin activities from the datasets. To define set of locations,
we look into how many unique places have this user checked in. For each pair of subsequent locations, we calculate the shortest route applying formula (\ref{haversine}). For the probability in the Markov Model of moving from location $a$ to location $b$, we calculate how many times the user checks in at location $a$ immediately after checking in at location $b$ divided by the number of times the user checks in at the location $a$. Finally, we calculate the time it takes for a given user to go from one checkin to another. The entire process is depicted in Fig. \ref{generating-mm}. 

After we have our empirical Markov Model built for each user\footnote{We use ``user" when referring to the dataset and ``node" when referring to the simulation. A node is built from the social network data provided by the users.}, we use Miller's coordinate projection to convert geographic space into a Cartesian coordinate system that preserve the triangle law of distances. Finally for mobility simulation, each node randomly gets assigned to one of its checkins. Then each node randomly picks with the assigned probability the location of the next checkin and moves directly to it using a straight line trajectory. Once the node reaches the new checkin, it repeats the process until the end of the simulation.  

Hence, the difference between the RWP mobility model and our FMM is that in the latter the space of travel is limited to the area of the checkins 
for each individual node. Moreover, each node moves differently based on its training set of checkins. For instance, an 
adult might be inclined to check in at work more often than a student. Finally, we have control over the frequency of encounters by selecting users (friends or non-friends) who live near from each other $\bar{D}(i,j) \approx 0$ or far away $\bar{D}(i,j) >> 0$ from each other. 


Given checkin points, we like to learn the following parameters: distance, affinity, and time. Distance refers to how far a user travels 
and maximum distance is the longest distance between any two checkins of this user. Affinity or frequency  
refers to how often a user checks into the same location. Time refers to the timestamp of the checkin, which is used to infer how fast 
a user moves from one location to the next given the distance between them and two subsequent checkins at those locations. The time of the checkin is also used for calculating affinity. The notation used in the following is summarized in Table \ref{notation-summary}.
		
\newcounter{rownum}
\setcounter{rownum}{0}
	\begin{table}[!htbp]\footnotesize
		\caption{Summary of Notations}
		\label{notation-summary}
		\begin{center}
    		\begin{tabular}{cll}
    		\hline 
    		\addtocounter{rownum}{1}\arabic{rownum}. & $\mathcal{N}$ & Set of nodes. \\
    		\addtocounter{rownum}{1}\arabic{rownum}. & $\mathpzc{t}$ & Mobility generation time \\
    		\addtocounter{rownum}{1}\arabic{rownum}. & $\mathpzc{x}$ & Width of field in $\mathbb{R}$ \\
    		\addtocounter{rownum}{1}\arabic{rownum}. & $\mathpzc{y}$ & Height of field in $\mathbb{R}$ \\
    		\addtocounter{rownum}{1}\arabic{rownum}. & $\alpha$ & Position tuple of latitude and longitude \\
    		\addtocounter{rownum}{1}\arabic{rownum}. & $\mathpzc{l}$ & Checkin Tuple of latitude, longitude, and timestamp \\
    		\addtocounter{rownum}{1}\arabic{rownum}. & $\mathcal{C}_{i}$ & Set of checkins of user $i$ \\ 
    		\addtocounter{rownum}{1}\arabic{rownum}. & $\mathcal{D}(i)$ &   Distance Matrix of user $i$   \\
			\addtocounter{rownum}{1}\arabic{rownum}. & $\mathcal{A}(i)$ &   Affinity Matrix  of user $i$ \\
			\addtocounter{rownum}{1}\arabic{rownum}. & $\mathcal{T}(i)$ &	Temporality Matrix of user $i$  \\
			\addtocounter{rownum}{1}\arabic{rownum}. & $CS(i,j) $ &	Checkin Similarity of user $i$, $j$ \\
			\addtocounter{rownum}{1}\arabic{rownum}. & $d(i,j) $ &	Average Distance between user $i$, $j$ \\ 	
    		\hline
    	\end{tabular}
     \end{center}
	\end{table}
	
	The \emph{distance matrix} $\mathcal{D}$ of node $i$ denoted as $\mathcal{D}(i)$ is an $ k_{i} \times k_{i} $ symmetric matrix defined as $\mathcal{D} = [\mathpzc{c}_{i,j}]_{k_{i}xk_{i}}$ where $k_{i}$ is the number of checkins of node $i$ and $\mathpzc{c}_{m,m} = 0$ $m \le k_{i}$ Clearly, $\mathpzc{c}_{n,m} =  	
\mathpzc{c}_{m,n}$, since the distance going from point $a$ to point $b$ is the same as going from point $b$ to point $a$. Last but not least, the distance from going from point $a$ to itself is 0, denoted as $\mathpzc{c}_{m,m} = 0$. 
		
	Hence, the average distance travelled by a user is defined as the average distance of all possible distances between any two 
	checkins. 
	
	\begin{equation}
		\frac{1}{k^{2}}\sum\limits_{m=1}^k\sum\limits_{n=1}^k \mathcal{D}[m,n] = \frac{2}{k(k - 1)}\sum\limits_{m=1}^k\sum\limits_{n=m+1}^k \mathcal{D}[m,n]
	\end{equation}
	
	Naturally, the \emph{affinity matrix} of node $i$ denoted as $\mathcal{A}(i)$ is an $k_{i} \times k_{i}$ matrix defined as
	$\mathcal{A}(i)=[\mathpzc{f}_{i,j}]_{k_{i}xk_{i}}$

	
	The $\mathpzc{f}_{m,n}$ is defined as follows

	\begin{equation}
		\mathpzc{f}_{m,n} = \frac{ \norm{\mathpzc{c}_{m} \rightarrow \mathpzc{c}_{n}}}
		{\norm{\mathpzc{c}_{m}}} 
	\end{equation}
	
		$\norm{\mathpzc{c}_{m} \rightarrow \mathpzc{c}_{n}}$ denotes the number of times checkin $c_{n}$ occurs immediately after checkin $c_{m}$ and  
		$\norm{\mathpzc{c}_{m}}$  denotes the number of times the location tuple $(c_{m}^{0},c_{m}^{1})$ appears as the location tuple of all checkins of user $i$. 
		
	The \emph{temporal matrix} of node $i$ denoted as $\mathcal{T}(i)$ is an $k_{i} \times k_{i}$ matrix 	
	defined as $\mathcal{T}(i) = [\mathpzc{t}_{m,n}]_{k_{i}xk_{i}} $
		
	where $\mathpzc{t}_{m,n} = 0 \iff m=n$, and $t_{m,n}$ is defined as follows 
	
	\begin{equation}
		t_{m,n} = \mathpzc{l}_{n}^{2} - \mathpzc{l}_{m}^{2}
	\end{equation}

	In other words, $t_{m,n}$ represents the time elapsed between checkin $m$ and checkin $n$. 

	Naturally, the maximum distance of travel for a given node is $\rho = max_{m\leq k_i,n\leq k_i}(c_{m,n})$. The average distance of travel is the average distance between the $k_{i}(k_{i}-1)$ checkins, and the 		
	maximum coverage of a given node is the area of the circle whose radius $\rho$ is able to connect any two checkins, which is 		
	$\pi\frac{\rho}{2}^2$. 
	Hence, the area of maximum coverage of a node is strictly greater than the maximum area
	generated by the given checkins, unless all the checkins lie on the perimeter of the circle. Using the area of the circle not only 	
	simplifies the calculations for finding the maximum coverage, but also allows some flexibility for noise within the friendship 	
	mobility model.

Since our dataset is limited, there is a chance that a user might enter an absorbing state of the MM (i.e, the state from which there is no transition out). Such a state will have zero probability of transition from itself. This is inconvenient in simulations, as we may enter this location in the middle of a simulation. To avoid such possibility, we can add an artificial probability of transition from such absorbing state to some random location. 

	Once we calculated the $\mathcal{D}\mathcal{A}\mathcal{T}$ of a subset of users in the dataset (in the case of our FMM, each subset is defined by a transitive friendship relation of a randomly chosen node), we can simulate the mobility 
	traces using the algorithm introduced in this section. 
	

\begin{figure*}[htp!]
  \subfigure[Friendship is bounded by distance.]{\includegraphics[width=60mm, height=40mm]{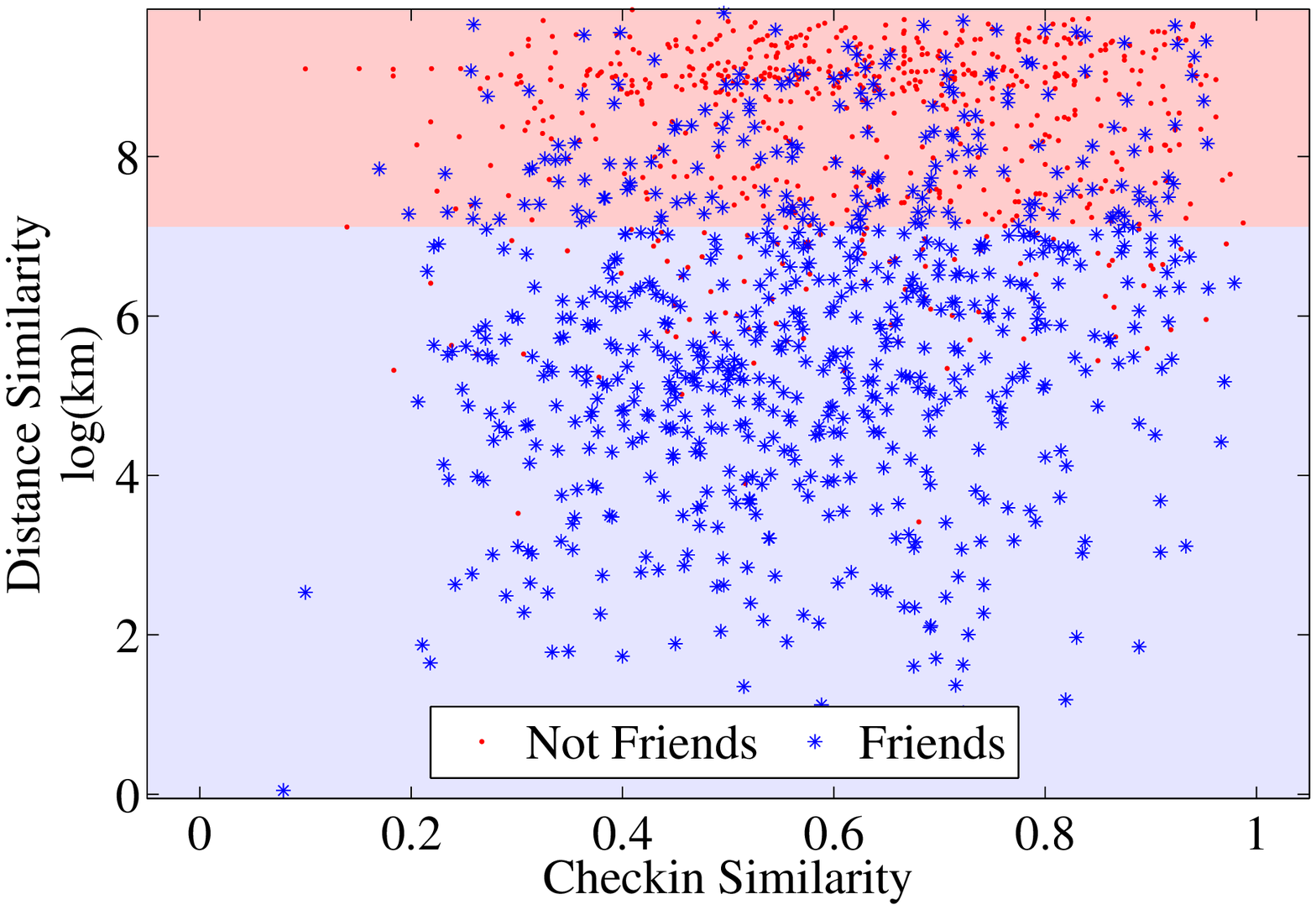}}
  \subfigure[Mobility traces of RWP and FMM.]{\includegraphics[width=60mm, height=40mm]{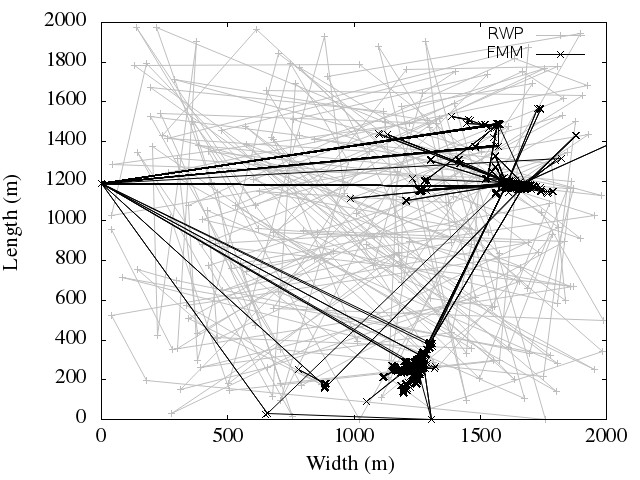}}
  \subfigure[Distance impacts friendship]{\includegraphics[width=60mm, height=40mm]{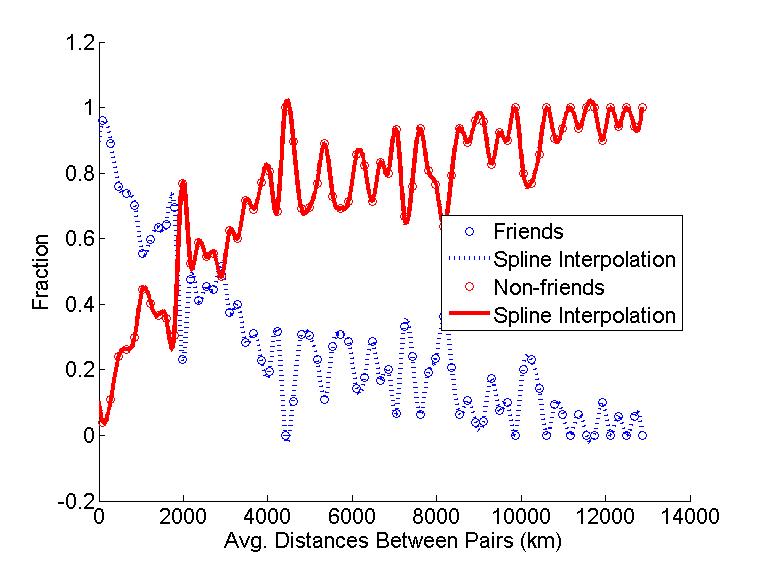}}
  \caption{}
  \label{subf1}
\end{figure*}

In Fig. \ref{subf1}(a), there are 701 blue points that represent two randomly selected users who are friends and 620 red points that represent 
two randomly selected users who are not friends within the dataset. The shaded region is drawn by using the k-nearest neighbor 
algorithm for classifying whether two users are friends given their average distance apart and checkin similarity. 

In Fig. \ref{subf1}(b), the gray lines represent the trajectories of the RWP. The black lines represent the trajectories of the FMM using the 
algorithm introduced in this paper for simulating human mobility. Notice how in our FMM, nodes are  allowed to move only to a subset 
of locations that replicate checkin behavior of humans versus random locations in the RWP.

In Fig. \ref{subf1}(c), the x-axis represents the average distance between two randomly selected users that could either by friends or non-
friends as defined in the social network. The y-axis represents the fraction of users who are friends represented by the blue line or 
non-friends represented by the red line. Roughly, 3000 randomly selected pairs were chosen for the class of friends and another 3000 
for the class of non-friends. 

\section{Protocols Evaluations}
\label{sec:protocol}
In the networking literature, the backoff timer in the MAC 802.11 is an algorithm implemented for preventing traffic congestion of wireless transmissions. If two transmissions are within radio range of each other and want to communicate through a wireless channel, one will randomly backoff to let the other one ``talk". Suppose we are interested in optimizing the performance of a wireless network at a conference where the attendees are working on their laptops and moving from location to location with some hidden attributes. Since humans do not move randomly, there will be more congestion at popular sessions. 

If we use the RWP model, the most congestion occurs in the middle due the stationary distribution. However, if we ask a random set of 
attendees to check into a particular room at the conference, we will know where the network congestion will be the highest. 

We designed a controlled experiment in MANET using ns-2 to compare the traffic congestion between the RWP and the FMM. In the experiment, there are 15 mobile nodes constantly sending out packets to their neighbours within the radio range. Other simulation parameters are listed in Table \ref{simulation-parameters}. When two or more nodes are within radio range of each other, at most one node can make a successful transfer and the remaining has to pause. We measure the overall congestion of the network by counting how many times did a node need to pause given that we know its current location duration the simulation. With the FMM, we were surprised that it had 2.77 times more congestion than the RWP. 

		\begin{table}[htbp]
  			\centering
  			\caption{Simulation Details}
    			\begin{tabular}{| l | c | c|}
    			\addlinespace
    				\hline
    			\textbf{Parameters}	& \textbf{RWP}	& \textbf{FMM} \\
    				\hline
    			Simulation Time ($\mathpzc{t}$) & 10,000s & 10,000s \\
       				\hline
       			MAC Layer & 802.11Ext & 802.11Ext \\
       				\hline
       			Width ($\mathpzc{x}$) & 2000m & 2000m \\
       				\hline
       			Length ($\mathpzc{l}$) & 2000m & 2000m \\
       				\hline
   				Nodes ($\mathpzc{n}$) & 15 & 15 \\
       				\hline
   				Pause Time & 0 & 0 \\
       				\hline
    			Min Speed & 0 & 5 \\	
       				\hline
    			Max Speed & 5 &  5\\
    				\hline
    			\emph{Total Backoffs}. & \emph{598,316} & \emph{1,654,967} \\
					\hline    
    				\bottomrule
    			\end{tabular}
  			\label{simulation-parameters}
		\end{table}

This agrees with our intuition that in the FMM, friends like to maintain their relationships by being closer to each other. Economic 
factors like the cost of transportation and mobility have a great impact on how we choose with whom to be friends.

\begin{figure*}[htb]
  \subfigure[Simulation Overview]{\includegraphics[width=60mm, height=40mm]{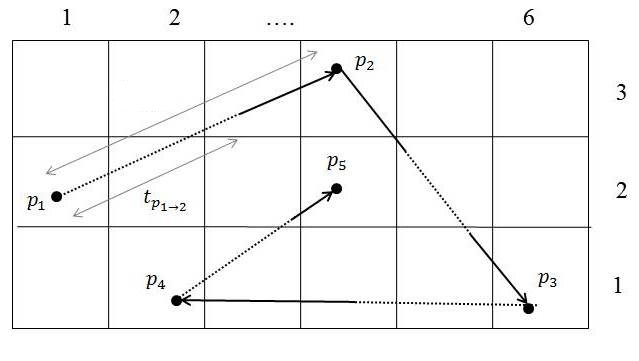}}
  \subfigure [Frequency of pauses using the RWP]{\includegraphics[width=60mm, height=40mm]{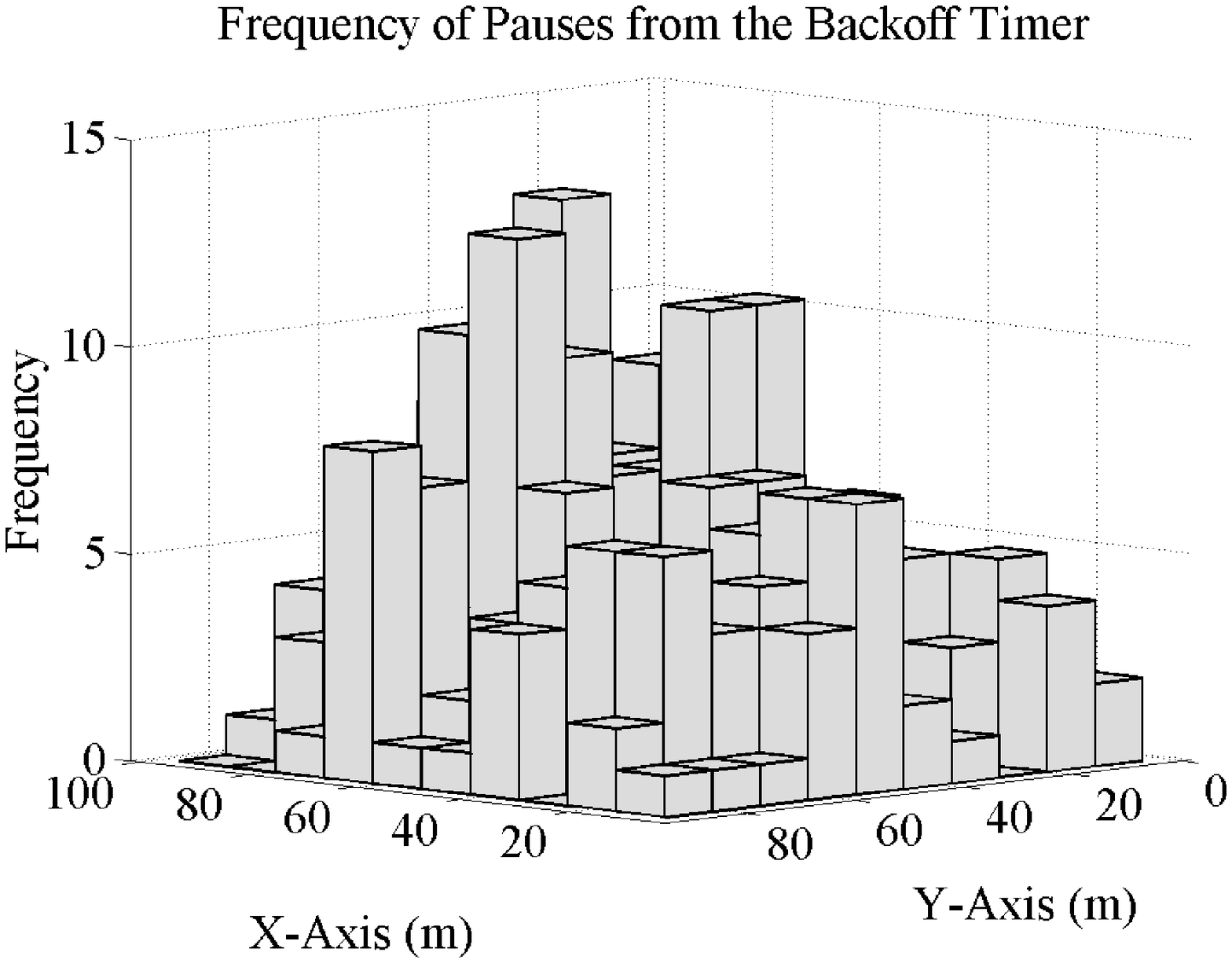}}
  \subfigure[Congestion in FMM and RWP]{\includegraphics[width=60mm, height=40mm]{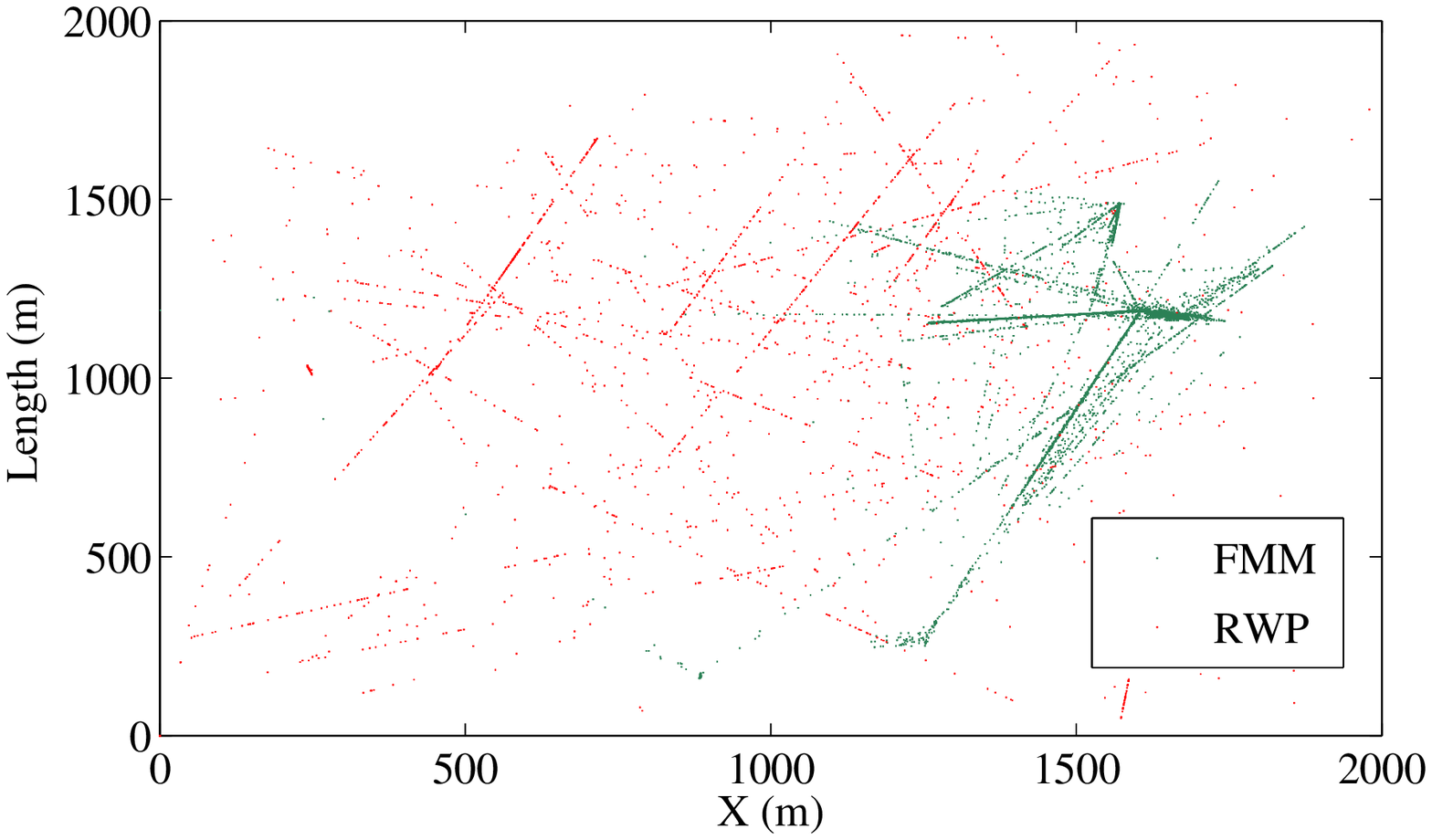}}
  \caption{}
  \label{subf2}
\end{figure*}

Fig. \ref{subf2}(a) provides the outline of a simulated node moving and how it causes congestion. Suppose a node starts at $p_{1}$ and travels to $p_{2}$ with some speed dictated by the mobility model. A mobile node cannot transmit 
if there is already a concurrent transmission within some nearby range. Therefore, it pauses until it detects no concurrent 
transmissions. The pause time duration in a subarea is the total amount of time of all the nodes pausing or suspending their 
transmissions due to the backoff timer of the MAC 802.11 protocol. During the trip from $p_{1}$ to $p_{2}$, the node pauses in 3 subareas (1,2), (2,2), (3,3) represented by the dashed line, meaning that the transmission was suspended for some time. The length of the dashed line in a subarea represents the duration of pause time for that particular trip.

Fig. \ref{subf2}(b) displays the frequency of pauses caused by the backoff timer in the MAC 802.11 protocol using the RWP. We noticed how congestion is centralized in the middle, which is correlated to the stationary distribution of the RWP. 

Fig. \ref{subf2}(c) displays the simulation results of network congestion in a controlled MANET. We took a sample of locations with traffic congestion. The points represent places where at least one node had to backoff within the simulation. Notice how traffic congestion is dispersed for RWP and clustered for FMM. Please note that this graph only shows places of congestion but not density or total volume of communications.

\section{Literature Review}
\label{sec:literature}

In \cite{human-mobility}, researchers used anonymized data from mobile phones to study individual mobility and concluded that human 
trajectories are predictable in a sense that time and space are occasionally repeated, like going from home to work and taking a 
vacation once in a while. In \cite{wang:mobility}, researchers investigated the interconnection of human mobility and social ties by 
examining mobile phone records with the goal of predicting links; i.e., given the mobile traces of the users, how to predict which new 
links will develop in the future. Using mobile datasets from cellular phones, they constructed a friendship network by looking at who-
call-whom in the phone call records. In \cite{cho:movement}, researchers looked at Gowalla and other datasets to examine how social 
relationships can be used to explain human mobility and to develop a model of human mobility by trying to fit behavior of checkins 
using Expectation-Maximization (EM). The main difference from our paper is that they want to predict the current location of a user 
given the time and day. In contrast, we only look at the transition of going from one location to another. 

Existing works have provided a comprehensive understanding of the MAC 802.11 from analytical and empirical perspectives using random 
processes and simulation results. In \cite{Razafindralambo_2006}, the authors have analytically studied four 
backoff algorithms with backoff suspension using multi-hop network scenarios. In \cite{Foh_2002}, the authors have 
provided an analysis of MAC 802.11 performance using a random process to determine the delay of transmission in a single-hop wireless 
network. Our empirical approach to measuring backoff relies on using location-based checkins to simulate node mobility. 
	
\section{Conclusion \& Future Work}
\label{sec:future-work}
\subsection{Conclusion}
Before closing, let's reiterate contributions of this paper. First, we took 
advantage of Gowalla, a recent and rising location-based social network, to answer questions about mobility as a function of social ties - how frequently do 
friends travel together and how does mobility impact friendship. By using Gowalla, we were able to collect traces of human mobility and topology of friendship. Combining 
these two elements together, we demonstrated that an important feature of traffic, network congestion, is dramatically affected by 
friendship mobility patterns. In a particular simulation scenario, network congestion increased by 94 percent when 
we replaced the RWP with our FMM.

Second, we discovered two interesting facts about friendship. In Fig. \ref{subf1}(a) and \ref{subf1}(c), we noticed that friendship decays almost exponentially as distance increases. However, the distance over which friendship is possible is large,
about 800 miles as shown in Fig. 3(a). We hypothesize that social media provide a mechanism for people to maintain friendship at such 
large distances. 

More interestingly, we also found that co-appearance represented by checking similarity is a poor indicator of friendship; that is, 
people who are temporarily within the same place and time are not likely to be friends. Co-appearance is not the same
as the average distance of separation for geo-proximity used in in Fig. 3(c).

Intuitively, co-appearance happens often at popular spots, like concerts and cafes that attract people living at great variety of locations. Even if a group of a few friends goes together for a concert, they would not be friends with thousands of other attendees, hence, a chance that a random pair of attendees are friends is low.  Our results confirm this intuition and indicate importance of personal face-to-face contacts for friendship. Occasional co-appearances are not sufficient, but geo-proximity helps in establishing and maintaining friendship, as shown by plot in Fig. 3(c).
 
Last, our FMM provides a more accurate and complex model of human mobility by taking into account of social ties. Such models are important for traffic engineering in communication networks as well as transportation systems, urban planning, and epidemiology. We believe that using FMM will improve accuracy of such applications and their results. 

\subsection{Future Work}
One interesting question relating to location-based social networking is whether social ties or economic factors influence human mobility. If human mobility is influenced by social or economic factors, then how to capture, explain, and verify this phenomenon is a difficult but worthy of study problem. 

We have observed that the patterns of checkins confirmed our intuition of humans making repetitive sequences of movements. For instance, consider a sequence of 
going from home to work, work to lunch, lunch back to work, and finally work to home. It cannot be captured by using a 
Markov Model (MM) because states are not dependent on just the previous state (e.g., the first departure from work is for lunch, while the second is for home). Therefore, it would be interesting to measure the effectiveness of different 
approaches for predicting the next checkin, such as using PCFG (probabilistic context free grammars) \cite{Geyik_2010}, or any other model richer than MM. Even though 
prediction and validation were not our objectives in this paper, in the future we would like to benchmark different models (HMM, PCGF, 
supervised learning, etc) to determine the accuracy of capturing the next location of a human being given the previous checkins as 
training data.

Finally, there have been advances in generating movements of dependent nodes in group mobility models. For instance, the Reference Point Group Mobility model (RPGM) \cite{WCM:WCM72} takes into consideration dependent nodes by putting them into groups or clusters. Each group has a center, which dictates the mobility of every node within the cluster. Many variants can be implemented within the concept of group mobility. Other interesting examples are City Selection Mobility Model, where the area in the mobility space is taken from a street within a city and Nomadic Community Mobility Model, where a group of nodes move together from one place to another location \cite{WCM:WCM72}. 

\section{Acknowledgement} 
Research was sponsored by the Army Research Laboratory and was
accomplished under Cooperative Agreement Number W911NF-09-2-0053. The
views and conclusions contained in this document are those of the
authors and should not be interpreted as representing the official
policies, either expressed or implied, of the Army Research Laboratory
or the U.S. Government. The U.S. Government is authorized to reproduce
and distribute reprints for Government purposes notwithstanding any
copyright notation here on.

\IEEEpeerreviewmaketitle
\bibliographystyle{plain}
\bibliography{mybib}
\nocite{*}
\end{document}